# Numerical and analytical research of the impact of decoherence on quantum circuits

Yu.I. Bogdanov[1,2,3], A.Yu. Chernyavskiy[1,4], B.I. Bantysh[1,2], V.F. Lukichev[1], A.A. Orlikovsky[1], I.A. Semenihin[1], D.V.Fastovets[1,2], A.S. Holevo[5]

1. Institute of Physics and Technology, Russian Academy of Sciences
2. National Research University of Electronic Technology MIET
3. National Research Nuclear University 'MEPHI'
4. Faculty of Computational Mathematics and Cybernetics, Moscow State University
5. Steklov Mathematical Institute, Russian Academy of Sciences

## ABSTRACT

Three different levels of noisy quantum schemes modeling are considered: vectors, density matrices and Choi-Jamiolkowski related states. The implementations for personal computers and supercomputers are described, and the corresponding results are shown. For the level of density matrices, we present the technique of the fixed rank approximation and show some analytical estimates of the fidelity level.

## 1. INTRODUCTION

The decoherence is the main obstacle to full-scale quantum information technologies. That is the reason why the modeling of noisy quantum schemes is very important task. We consider three different levels of such modeling: state-vectors, density matrices and Choi-Jamiolkowski relative states.

The modeling of ideal quantum circuits has been widely studied [1-5], while the noise can be easily incorporated by Monte Carlo technique [6]. A brief description of this approach, the features of realization, and some results are described in Section 2.

The third section is devoted to the density matrix level, which allow us to consider the noise in terms of Kraus operators [7-10]. We present the technique of fixed rank, which drastically increases the number of modeled qubits. This technique is analyzed on the Grover algorithm and quantum Fourier transform. Moreover, we show that the rank-1 approximation leads to accurate analytical estimates of fidelity.

The forth section is devoted to the modeling in terms of Choi-Jamiolkowski relative states [9-11]. This approach allows us to analyze the gates and circuits without particular reference to the input state. We describe the implemented software package for the noisy circuits simulation and show the results of its usage for simple error correction schemes and QFT.

## 2. THE LEVEL OF STATE-VECTORS

In this case, we have an initial state $|\psi_0\rangle$ and the set of unitary operations $\{U_i\}$. The action of the circuit is defined by sequential multiplications of unitary operators: $U_s \cdot U_{s-1} \cdot ... \cdot U_2 \cdot U_1 |\psi_0\rangle$. Operations $U_i$ usually affect one, two

*e-mail: bogdanov_yurii@inbox.ru

or sometimes three qubits. The one-qubit operation can be represented as $b_{i_1...i_k...i_n} = u_{i_k 0} \cdot a_{i_1...0...i_n} + u_{i_k 1} \cdot a_{i_1...1...i_n}$, where $a_{i_1 i_2...i_n}$ and $b_{i_1 i_2...i_n}$ are the original and the new amplitudes respectively, and the equation for the multi-qubits case can be easily derived in a similar way. The non-parallel realization of such transformations is trivial, but it is clear that due to the arbitrary memory dependence during modeling the parallel version of such modeling is much more difficult. However, such algorithm can be made more efficient using the active messages approach. Such approach was implemented to model 39-qubit schemes using the "Lomonosov" supercomputer [3], as well as MVS-100K and MVS-10P supercomputers [6]. We notice that using the active messages programming paradigm leads to an approximately 10 times speed-up.

On this level the noise can be used in Monte Carlo approach: every gate is replaced by its stochastic noisy version. For example, an arbitrary one-qubit gate $U$ can be replaced by $U = U_{ideal} \cdot V_{noise}$, where

$$V_{noise} = \begin{pmatrix} \cos(e\xi) & \sin(e\xi) \\ -\sin(e\xi) & \cos(e\xi) \end{pmatrix}, \quad (1)$$

where $\xi$ is normally distributed by $N(0,1)$, $e$ is an error rate.
The controlled phase-shift gate

$$U_\theta = \begin{pmatrix} 1 & 0 & 0 & 0 \\ 0 & 1 & 0 & 0 \\ 0 & 0 & 1 & 0 \\ 0 & 0 & 0 & e^{-i\theta} \end{pmatrix}, \quad (2)$$

which is used in quantum Fourier transform, and can be represented as follow: the only non-unitary diagonally element of the matrix $e^{-i\theta}$ is being replaced by $e^{-i(\theta + e \cdot \xi)}$.

The results of Monte-Carlo modeling of Grover's algorithm are presented on Fig.1.

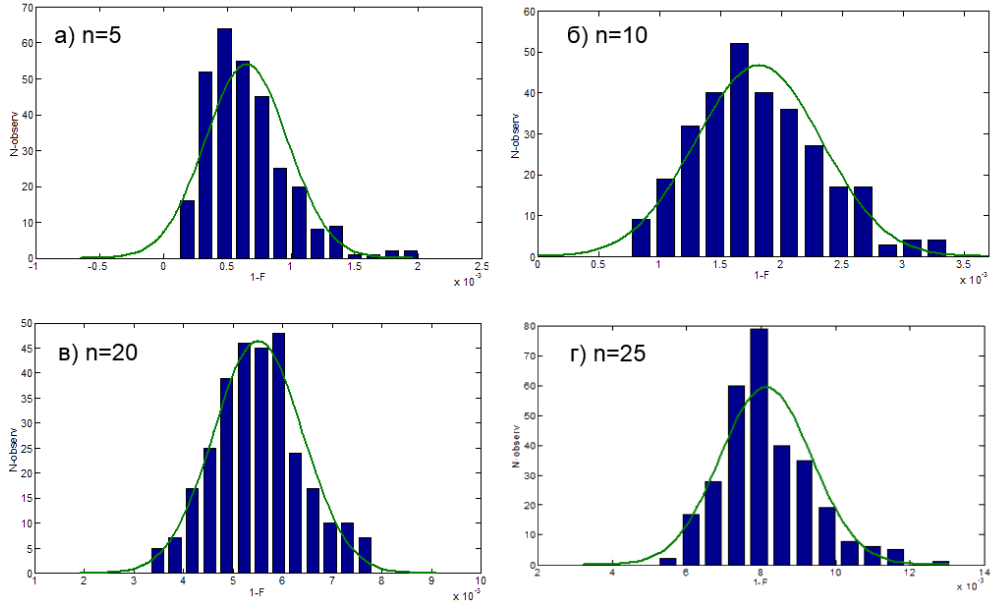

Fig.1. The distribution of the error (1-Fidelity) of Grover's algorithm with different number of qubits.

# 3. THE LEVEL OF DENSITY MATRICES AND FIXED RANK APPROXIMATION

Now we have an initial density matrix $\rho_0$, and unitary operations transforms $\rho$ to $U\rho U^\dagger$. This operation can be reduced to the vector case. The arbitrary matrix $\rho$ can be written in a ket-bra form: $\rho = \sum_{i,j} \rho_{ij} |i\rangle\langle j|$, by conjugating the conjugated space, we obtain an isomorph square-dimensional vector $|\psi_p\rangle = \sum \rho_{ij} |i\rangle|j\rangle$, and the unitary transformation becomes $U\rho U^\dagger \mapsto U \otimes U|\psi\rangle$. So, for example, if we can simulate two-qubit operations on vectors, we can do one-qubit operations on density matrices.

Moreover, we can move from Monte Carlo noise representation to quantum operations. Arbitrary quantum operation can be represented using Kraus operators:

$$\mathcal{E}(\rho) = \sum_i E_i \rho E_i^\dagger, \qquad (3)$$

where operators $E_i$ are called Kraus operators and $\sum_i E_i^\dagger E_i = I$.

The modeling of arbitrary scheme with gates represented by Kraus operators was implemented on "Lomonosov" supercomputer and represented for up to 17 qubits.

## 3.1 Kraus operators representation of Hadamar and phase shift gates with Monte Carlo noise

For the further analysis we need to find Kraus operators of noisy versions of used gates.

Ideal Hadamar operator in matrix form

$$H = \frac{1}{\sqrt{2}} \begin{pmatrix} 1 & 1 \\ 1 & -1 \end{pmatrix}, \qquad (4)$$

Noisy version of this gate is obtained by multiplication ideal operator $H$ and Monte Carlo noise operator

$$H_e = H \cdot V_{noise}, \qquad (5)$$

Monte Carlo noise of Hadamar operator which correspond noise matrix $V_{noise}$ in (1) has Kraus operators

$$E_1 = \sqrt{\lambda_1} \begin{pmatrix} 1 & 0 \\ 0 & 1 \end{pmatrix}, \quad E_2 = \sqrt{\lambda_2} \begin{pmatrix} 0 & 1 \\ -1 & 0 \end{pmatrix}, \qquad (6)$$

where

$$\lambda_1 = \frac{1}{2}(1 + \exp(-2e^2)), \quad \lambda_2 = \frac{1}{2}(1 - \exp(-2e^2)).$$

The controlled phase-shift gate (2) we replaced by its noisy version

$$U_e = U_\theta \cdot U_R, \qquad (7)$$

where

$$U_R = \begin{pmatrix} 1 & 0 & 0 & 0 \\ 0 & 1 & 0 & 0 \\ 0 & 0 & 1 & 0 \\ 0 & 0 & 0 & e^{-ie\xi} \end{pmatrix}, \qquad (8)$$

we can represent this noise operator by Kraus operators:

$$E_1 = \begin{pmatrix} 1 & 0 & 0 & 0 \\ 0 & 1 & 0 & 0 \\ 0 & 0 & 1 & 0 \\ 0 & 0 & 0 & \sqrt{P} \end{pmatrix}, \quad E_2 = \begin{pmatrix} 0 & 0 & 0 & 0 \\ 0 & 0 & 0 & 0 \\ 0 & 0 & 0 & 0 \\ 0 & 0 & 0 & \sqrt{1-P} \end{pmatrix}, \quad (9)$$

where $P = \exp(-e^2)$.

### 3.2 Fixed rank approximation of simulation

The simulation always starts from pure (rank-1) density matrix initial state, and every gate can increase the rank not more than in number of Kraus operators times. After every step (one gate) we can approximate our matrix by the matrix of the fixed rand using spectral decomposition. Such approximation violates the trace norm and we need to normalize the result to obtain the correct density matrix (positivity is obviously not violated). Fig.2 shows the accuracy of such approximation for the calculation of fidelity during Grover's algorithm.

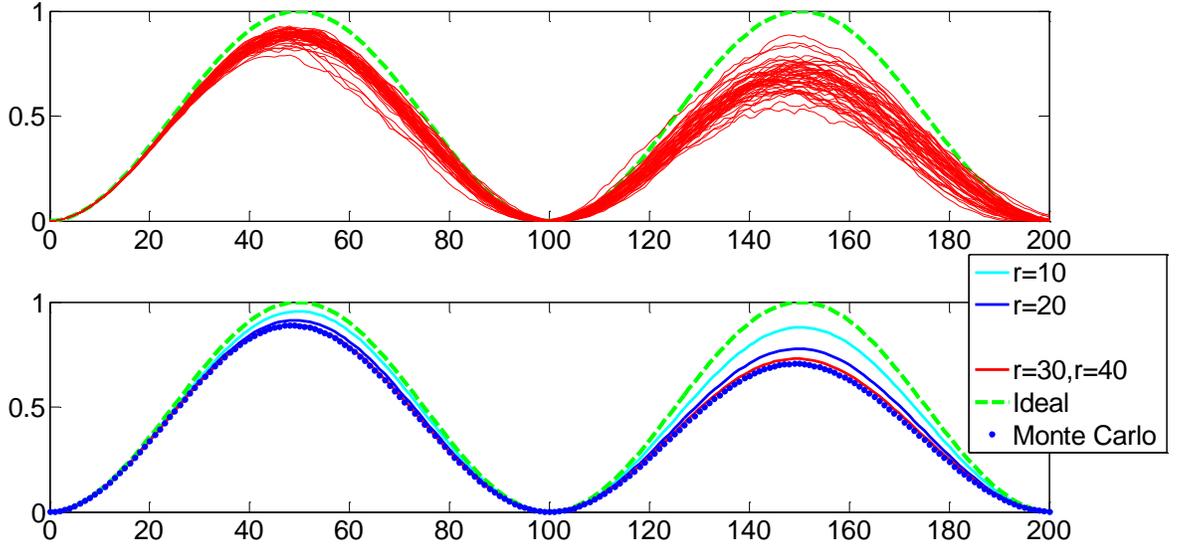

Fig. 2. Dynamics of fidelity to target state in noisy Grover's algorithm (12 qubits, e=0.01). Upper: different iterations of Monte Carlo noise modeling; Lower: fixed-rank approximation (normalized) and averaged Monte Carlo.

The normalization of the lower-rank matrix is needed for the formal accuracy, but it turns out, that the the fidelity unnormalized matrix is much more precise (Fig.3). We notice, that this is true only for the special cases like Grover's algorithm (it will be shown further).

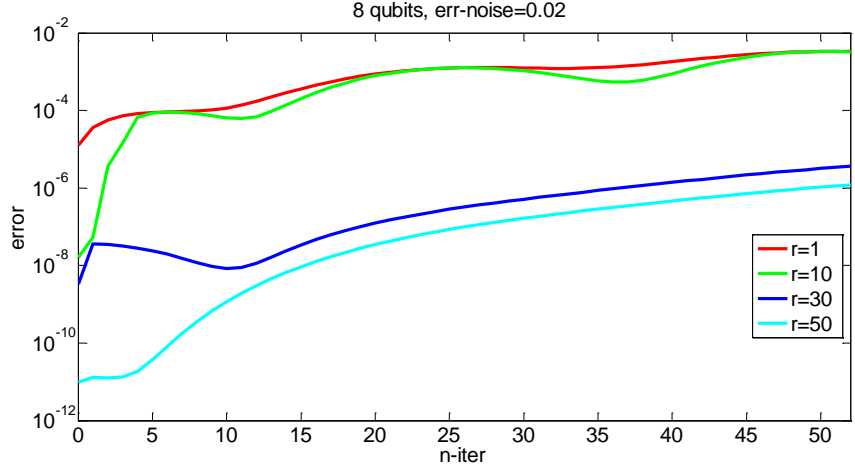

Fig. 3. The dependence of the error of fidelity approximation (unnormalized version) on the matrix rank.

### 3.3 Analytical estimates of fidelity by rank-1 approximation

As we can see on Fig.4 even the rank-1 unnormalized approximation shows the good accuracy. Therefore, we can obtain the fidelity (the probability of obtaining the correct answer) of Grover's algorithm analytically:

$$p_{noise}(j) = \lambda_1^{(n+2nj)} p_{ideal}(j), \quad (10)$$

where $j$ is the step of the algorithm, $p_{noise}$ and $p_{ideal}$ are the probabilities of obtaining the correct answer by the noisy and ideal algorithm respectively, $n$ is the number of qubits, $\lambda_1 = \frac{1}{2}(1 + \exp(-2e^2))$. The $n$ in the power corresponds to the noise of initial state preparation and $2nj$ corresponds to two Hadamar gates on each step and qubit.

For the QFT circuit the calculation of a good analytical estimate of fidelity is not so easy. By using the direct unnormalized rank-1 approach we can obtain simple lower bound for the fidelity of the whole circuit:

$$F_{QFT} = P_H^n P_R^{n(n-1)/8}, \quad (11)$$

where $P_H = \frac{1}{2}(1 + \exp(-2e^2))$, $P_R = \exp(-e^2)$ are the contributions of Hadamar and phase-shift gates respectively.

We can see on Fig. 4 that such estimate is rough and we need to improve it. It can be done by some kind of "main components" reduction of Kraus operators of the phase shift gate (7).

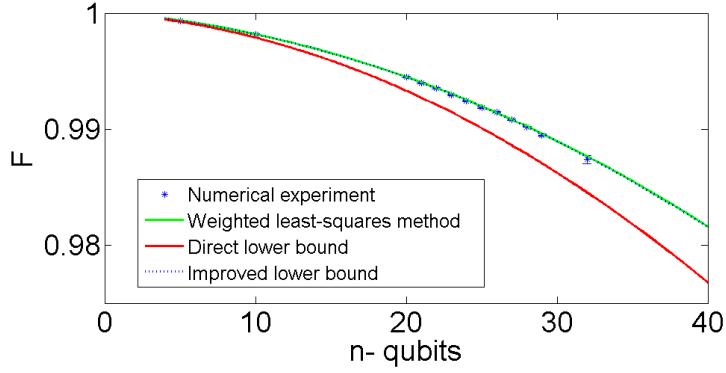

Fig.4. Fidelity of QFT circuit (e=0.01) and its analytical lower bound calculated by the direct and improved rank-1 approximation.

As it's well known Kraus operators are ambiguous due to some unitary transformation $U = \{u_{ij}\}$:
$E_i$ produces the same quantum operation as $B_i = \sum_j u_{ij} E_j$.

Kraus operators of the controlled phase shift (7) are diagonal, so we will find the SVD decomposition of the matrix, which consist their diagonal:

$$\psi = \begin{pmatrix} 1 & 1 & 1 & a \\ 0 & 0 & 0 & b \end{pmatrix}, \qquad (12)$$

where $a = \sqrt{P}$, $b = \sqrt{1-P}$.

The SVD decomposition is $\psi = USV^\dagger$, where

$$S = \begin{pmatrix} \sqrt{\lambda_1} & 0 \\ 0 & \sqrt{\lambda_1} \end{pmatrix},$$

$\lambda_1 = 2 + \sqrt{4 - 3b^2} = 2 + \sqrt{1 + 3P}$,

$\lambda_2 = 2 - \sqrt{4 - 3b^2} = 2 - \sqrt{1 + 3P}$.

The unitary matrix is

$$U = \begin{pmatrix} \dfrac{1}{\sqrt{1+f^2}} & \dfrac{-f}{\sqrt{1+f^2}} \\ \dfrac{f}{\sqrt{1+f^2}} & \dfrac{1}{\sqrt{1+f^2}} \end{pmatrix}, \qquad (13)$$

where

$$f = \frac{\sqrt{4 - 3b^2} - a^2 - 1}{ab} = \frac{\sqrt{1 + 3P} - P - 1}{\sqrt{P(1-P)}}.$$

This unitary matrix provides the necessary transformation of Kraus operators.

New Kraus operators are

$$\widetilde{E_1} = \frac{1}{\sqrt{1+f^2}} \begin{pmatrix} 1 & 0 & 0 & 0 \\ 0 & 1 & 0 & 0 \\ 0 & 0 & 1 & 0 \\ 0 & 0 & 0 & \sqrt{P}+f\sqrt{1-P} \end{pmatrix}, \tag{14a}$$

$$\widetilde{E_2} = \frac{1}{\sqrt{1+f^2}} \begin{pmatrix} -f & 0 & 0 & 0 \\ 0 & -f & 0 & 0 \\ 0 & 0 & -f & 0 \\ 0 & 0 & 0 & -f\sqrt{P}+\sqrt{1-P} \end{pmatrix}, \tag{14b}$$

where, again, $P = \exp(-e^2)$, $f = \frac{\sqrt{1+3P}-P-1}{\sqrt{P(1-P)}}$.

The improved lower bound is:

$$F_{QFT} = P_H^n \tilde{P}_R^{n(n-1)/8}, \tag{15}$$

where

$$P_H = \frac{1}{2}(1+\exp(-2e^2)), \quad \tilde{P}_R = \frac{(\sqrt{P}+f\sqrt{1-P})^2}{(1+f^2)^4}.$$

The Fig. 4 shows that the improved estimate has much better precision and visually agree with numerical experiment.

### 4. THE LEVEL OF CHOI-JAMIOLKOWSKI RELATIVE STATES.

An arbitrary quantum operation can be represented by the corresponding density matrix of the larger dimension. Such isomorphism is called the Choi-Jamiolkowski isomorphism [9-11]. To construct the Choi-Jamiolkowski relative state for the $d$-dimensional operation we need to take a maximally-entangled state

$$\frac{1}{\sqrt{d}} \sum_{i=1}^{d} |i\rangle|i\rangle, \tag{16}$$

and apply the operation to the first subsystem:

$$\rho_\mathcal{E} = \frac{1}{d} \sum_{i,j=1}^{d} \mathcal{E}(|i\rangle\langle j|) \otimes |i\rangle\langle j|. \tag{17}$$

Because of Choi-Jamiolkowski isomorphism we can analyze the accuracy of a quantum circuit not only for a concrete input, but as a whole: we consider the fidelity between relative states of a noisy and ideal circuit.

The ideas of the modeling of noisy gates and circuits can be found in [10]. The Matlab software package for the simulation of quantum circuits under amplitude (T1) and phase (T2) relaxation was implemented. It can compute the Choi-Jamiolkowski relative state for arbitrary circuits of 7 qubits. For the circuits with small amount of gates it can proceed more qubits because of the sparse structure of a relative state.

The described package was used to calculate the accuracy of simple error-correction codes and quantum Fourier transform.

**4.1 Three-qubits correction code**

This quantum code corrects a single classic error (Fig. 5) [8]. If we add Hadamard gates in circuit then code will correct phase error. The phase error correction code has been simulated as a test sample.

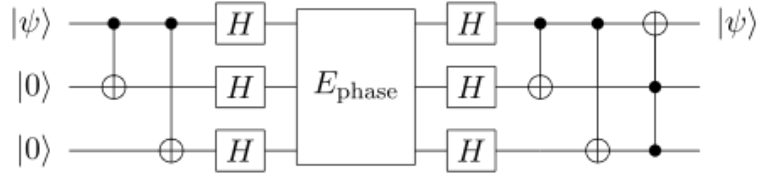

Fig. 5. Three-qubits correction code scheme.

If $p$ is the probality of phase error in a single qubit, then the minimum estimate of fidelity is

$$F_{min} = \sqrt{1 - 3p^2 + 2p^3}, \qquad (18)$$

The presence of the phase error is equal to phase relaxation procees, and the probability of this error is associated with $T_2$ (parameter of phase relaxation) by the dependency

$$T_2 = -\frac{t}{\ln(1 - 2p)}; \qquad (19)$$

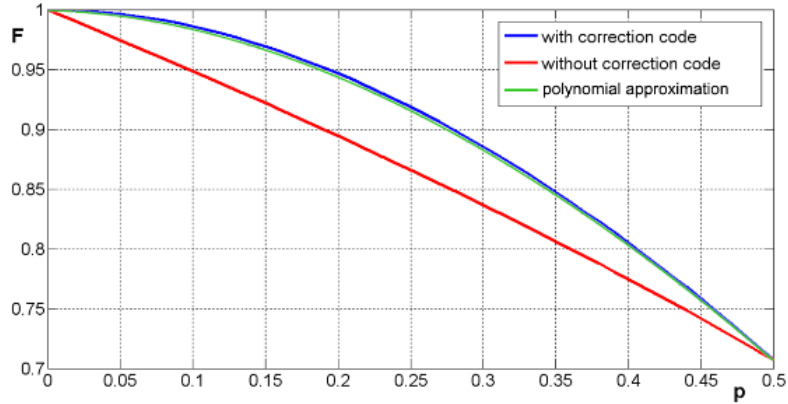

Fig. 6. Dependence of fidelity from phase error probability.

The polynomial approximation of the calculated dependence (Fig. 6) shows the excellent agreement with theoretical estimates.

### 4.2 Shor's code

This 9-qubit correction code correct an arbitrary in a single qubit.[8,12] The scheme of this code is based on three-qubits correction code.

Consider the scheme, taking into account only the phase error ($p$ - probality of phase error in single qubit). All qubits undergoes phase relaxation. Fig. 7 shoes the efficiency of Shor's code in the area of small values of phase error probability.

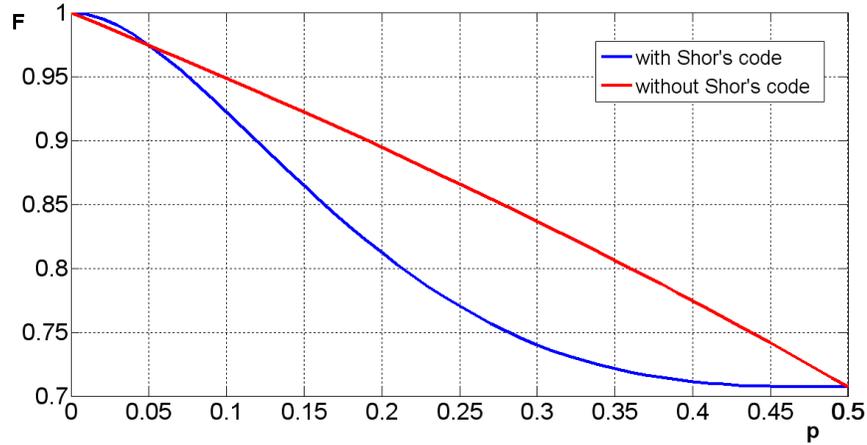

Fig. 7. Dependence of fidelity from phase error probability.

Fig. 8 shoes the influence of amplitude and phase relaxation on Z-gate corrected by Shor's code.

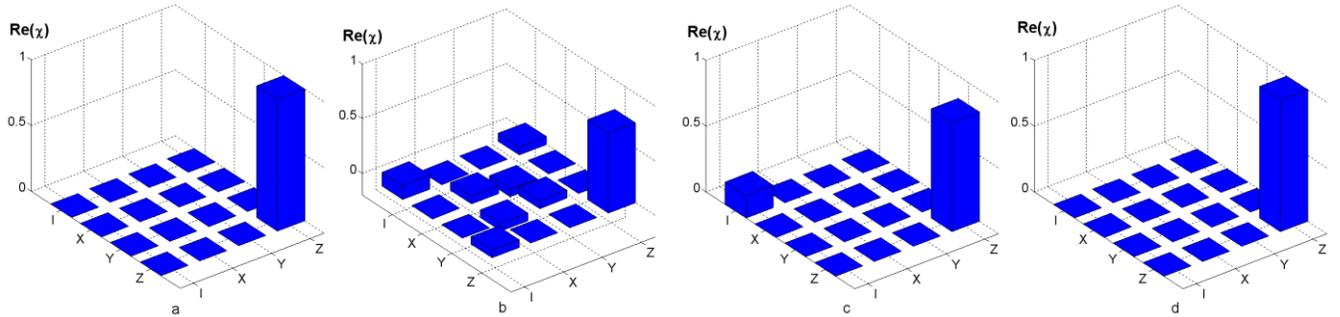

Fig. 8. Choi-Jamiolkowski relative states of gate Z. *a* – ideal state; *b* – noisy state ($T_1 = 3$, $T_2 = 2$); *c* - noisy state ($T_1 = 3$, $T_2 = 2$) with using three-qubits correction code; *d* - noisy state ($T_1 = 3$, $T_2 = 2$) with using Shor's correction code.

### 4.3 Quantum Fourier transform

As a more complex example, we took the 5-qubit quantum Fourier transform. Fig.9 shows the influence of T1 and T2 parametetrs on the fidelity of the QFT circuit.

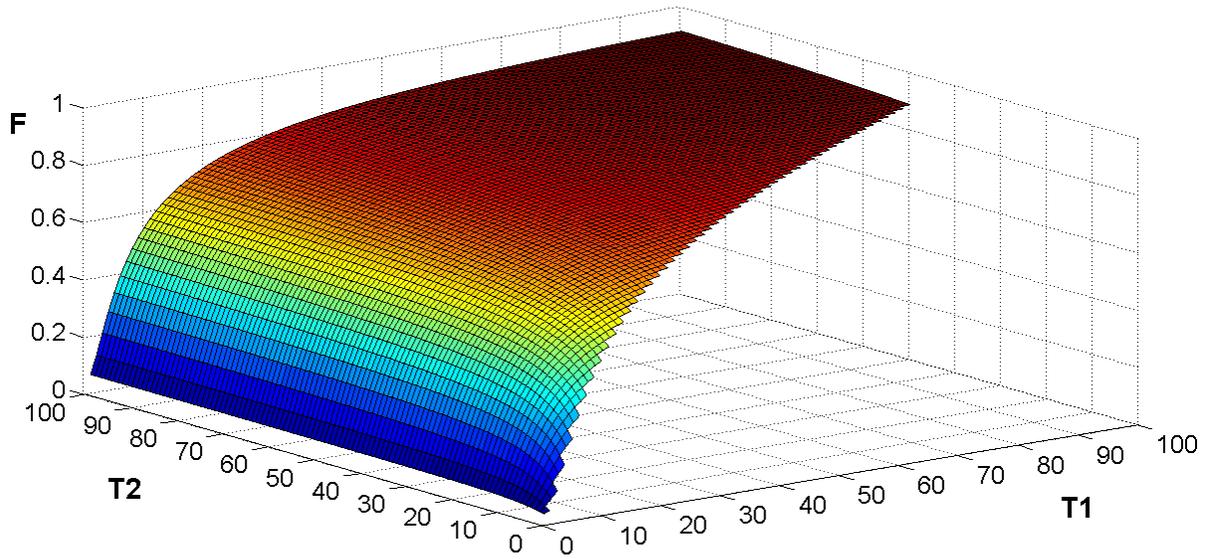

Fig. 9. Dependence of fidelity of QFT on parameters of amplitude and phase relaxation ($T_1$ and $T_2$ respectively).

## 5. CONCLUSIONS

Three levels of noisy quantum circuits modeling were presented. The level of state-vectors allows to analyze many-qubit schemes: the impact of noise on the quality of quantum information systems was analyzed for up to 39 qubits. The density matrix level allows taking the quantum operations noise models into account, and the fixed rank approximation of this approach leads to the precise estimates of the fidelity of Grover's algorithm and the Quantum Fourier transform. The Choi-Jamiolkowski relative state modeling allows to analyze quantum circuits as a whole, independently from input states: such analysis was made for QFT and error correction codes. The developed software packages are the initial step on the way of creating of the quantum information CAD system. Such system must be oriented on the automatic projecting of fault-tolerant quantum information technologies and the analysis of their quality and reliability.

## ACKNOWLEDGEMENTS

The authors thank the Joint Supercomputer Center of the Russian Academy of Sciences (JSCC RAS) for providing computational facilities.

This work was supported in part by Russian Foundation of Basic Research (projects, 13-07-00711, 14-01-00557), and by the Program of the Russian Academy of Sciences in fundamental research.